\documentclass[conference]{IEEEtran}
\usepackage{cite}
\usepackage{amsmath,amssymb,amsfonts}
\usepackage{algorithmic}
\usepackage{graphicx}
\usepackage{textcomp}
\usepackage{xcolor}
\usepackage{url}
\def\BibTeX{{\rm B\kern-.05em{\sc i\kern-.025em b}\kern-.08em
    T\kern-.1667em\lower.7ex\hbox{E}\kern-.125emX}}
\begin{document}

\title{User Centric Evaluation of Code Generation Tools
}
\IEEEspecialpapernotice{(Invited Paper)}

\author{\IEEEauthorblockN{Tanha Miah and Hong Zhu}
\IEEEauthorblockA{\textit{School of Engineering, Computing and Mathematics,}
\textit{Oxford Brookes University}\\
Oxford OX33 1HX, UK. 
Email: hzhu@brookes.ac.uk}
}

\maketitle

\begin{abstract}
With the rapid advance of machine learning (ML) technology, large language models (LLMs) are increasingly explored as an intelligent tool to generate program code from natural language specifications. However, existing evaluations of LLMs have focused on their capabilities in comparison with humans. It is desirable to evaluate their usability when deciding on whether to use a LLM in software production. This paper proposes a user centric method for this purpose. It includes metadata in the test cases of a benchmark to describe their usages, conducts testing in a multi-attempt process that mimics the uses of LLMs, measures LLM generated solutions on a set of quality attributes that reflect usability, and evaluates the performance based on user experiences in the uses of LLMs as a tool. 

The paper also reports a case study with the method in the evaluation of ChatGPT's usability as a code generation tool for the R programming language. 
Our experiments demonstrated that ChatGPT is highly useful for generating R program code although it may fail on hard programming tasks. The user experiences are good with overall average number of attempts being 1.61 and the average time of completion being 47.02 seconds. Our experiments also found that the weakest aspect of usability is conciseness, which has a score of 3.80 out of 5. 
\end{abstract}


\section{Introduction}
The past few years have seen a rapid growth of machine learning (ML) techniques in the development of large language models (LLMs).  The most advanced LLMs like ChatGPT\footnote{\url{https://openai.com/chatgpt}} and Gemini\footnote{\url{https://deepmind.google/technologies/gemini}} have demonstrated their impressive performance in completing natural language processing (NLP) tasks \cite{r1}. Although their primary goal is for NLP, they are also capable of performing various programming tasks with input in natural language and/or partially completed program code, because some of them have included program codes in their training dataset in addition to large volumes of natural language texts. Typical examples of such LLMs include OpenAI's GPT-3.5 underlying ChatGPT and Google's LaMDA model underlying Bard. Another approach to achieve the capability of performing coding tasks is through fine tuning of a pretrained LLM with program code datasets to build special purpose intelligent programming tools. Well-known examples of such systems include the Codex model underlying GitHub's Copilot. Many ML models have also been purposely built for various programming tasks. For example, AlphaCode \cite{r2}, InCoder \cite{r3}, StarCoder \cite{r4}, PolyCoder \cite{r5}, Replit's Ghostwrite\footnote{\url{https://replit.com/}} and Tabnine\footnote{\url{https://www.tabnine.com/}} have been developed for code completion, code generation, code comment summary generation, and debugging (code correction) tasks. ChatUNitTest \cite{r6}, TestPilot \cite{r7}, etc. are for the generation of test codes; see \cite{r8} for a survey. It is widely believed that it is becoming practical to automatically generate program code from natural language inputs; see, for example, \cite{r9, r10, r11}. 

However, a research question remains whether such ML based intelligent programming tools are useful for software development. Existing testing and evaluation of LLMs have focused on their capability in comparison with human intelligence. The evaluation results hardly represent LLM's usability from a user's point of view as a software development tool. Meanwhile, evaluations of LLMs reveal that LLMs like ChatGPT \emph{``is still far from achieving the ability to reliably solve many challenging tasks.''} \cite{r12 ,  r13} Moreover, they seriously suffer from the so-called hallucination problem, i.e., producing plausible sounding but incorrect or nonsensical answers \cite{r14}. 

Therefore, it is highly desirable to systematically test and evaluate LLMs' usability as an intelligent tool for generating program code. 
In this paper, we propose a user centric approach to the testing and evaluation of LLM's usability as code generation tools. Its applicability is demonstrated in a case study on testing and evaluation of ChatGPT's usability for generating program code in the R language from natural language input of functional specifications. 

The paper is organized as follows. Section \ref{sec:RelatedWork} reviews related works and discusses the problems associated with the existing methods of testing and evaluation of LLMs' capability of coding. Section \ref{sec:ProposedMethodology} presents the proposed user centric methodology. Section \ref{sec:CaseStudy} and \ref{sec:ExperimentResults} report a case study with ChatGPT to generate R program code.  Section \ref{sec:Comparison} compares the proposed method with the existing works. Section \ref{sec:Conclusion} concludes the paper with a summary and a discussion of the directions for future work. 

\section{Related Work}\label{sec:RelatedWork}

Evaluation of ML model's performance is indispensable to ML research and application. A great amount of work has been reported in the literature. Here we focus on related work at two levels: the evaluations of LLMs in general and the evaluations of LLMs' capability of code generation from natural language functional specifications.

\subsection{General Evaluation of LLMs}

The announcement of a new ML model or the proposal of a new ML technique commonly accompanied by a report on the evaluation of its performance. Such evaluations typically use common datasets to compare ML models to demonstrate one model's superiority over others. This evaluation method is called \emph{benchmarking}, and the datasets are called \emph{benchmark datasets}, or shortly \emph{benchmarks}. 

For example, the recent launch of Gemini \cite{r1} reports the performances of two versions of the ML model, i.e. Gemini-Ultra and Gemini-Pro, on many tasks of multi-modal language processing, which include multi-task language understanding, general reasoning, mathematical reasoning, and program code generation. It is compared with many front runners of LLMs, such as GPT-4, GPT-3.5, PalM 2-L, Inflection-2, Grok 1 and LLAMA-2. Table \ref{tab0} lists some commonly used text datasets to benchmark LLMs on natural language text processing tasks. They are also used by the Gemini team \cite{r1}. 
\begin{table}[htbp]
\caption{Commonly Uses Benchmarks for NLP.}
\begin{center}
\begin{scriptsize}
\begin{tabular}{|l|p{5.5cm}|}
\hline
\textbf{Name} & \textbf{Dataset}\\
\hline
\emph{MMLU} \cite{r15} &Multichoice questions in 57 professional and academic subjects. \\
\hline
\emph{GSM8K}\cite{r16} &Graduate school math questions.\\
\hline
\emph{MATH} \cite{r17} &Mathematics problems across 5 difficulty levels and 7 subdisciplines. \\
\hline
\emph{BIG-Bench-Hard}\cite{r18} &Hard Big-Bench tasks written as CoT problems. \\
\hline
\emph{HumanEval}\cite{r19} &Python coding tasks. \\
\hline
\emph{Natural2Code} \cite{r1} &Python code generation problems. \\
\hline
\emph{DROP}\cite{r20} &Reading comprehension and arithmetic problems. \\
\hline
\emph{HellaSwag}\cite{r21} &Common sense multiple-choice questions. \\
\hline
\emph{WMT23}\cite{r22} &Machine translation problems. \\
\hline
\end{tabular}
\end{scriptsize}
\label{tab0}
\end{center}
\end{table}

Independent evaluations of LLMs also follow the same paradigm of benchmarking. For example, Laskar et al. \cite{r13} evaluated ChatGPT on 140 tasks including code generation using the HumanEval and MBPP datasets \cite{r23}. 

It is worth noting that the evaluation of LLMs has all focused on their capability of performing tasks in comparison with human intelligence rather than their usability as a tool. 

\subsection{Evaluation of LLMs on Code Generation}

As discussed in the previous subsection, the capability of code generation is included in many of the general evaluations of LLMs. Much more work on the evaluation of LLMs' capability of code generation has also been reported in the literature with the development of intelligent coding tools. Several datasets have been constructed and corresponding evaluation methods advanced.

The APPS dataset developed by Hendrycks et al. in 2021 \cite{r24} is perhaps the first benchmark specifically constructed for evaluating code generation from natural language input as functional specification. Earlier works have been reported in the literature for other coding tasks. For example, in 2019, Kulal et al. evaluated the SPoC model, which generates code from pseudocode written in natural language \cite{r25}. In 2020, Lachaux et al. evaluated a ML model for code translation \cite{r26}. 

Strictly speaking, Kulal et al.'s SPoC model does not belong to code generation from natural language functional specifications. The input to SPoC consists of a pseudocode and a set of test cases. However, they set two important milestones on the method of evaluating ML model's performance on coding tasks. One is how to check the correctness of the generated code and the other is a metric to measure the performance of the ML model. These have become the common practice now. 

How to check the correctness of generated code automatically is an important issue. In the evaluations of NLP capabilities, the common approach is matching output with reference solutions either exactly or fuzzily, for example, using the ROUGE and BLEU metrics. However, for code generation, ROUGE and BLEU cannot capture the semantics of code. For example, Hendrycks et al. compared BLEU score with testing correctness \cite{r24}. They found that \emph{``BLEU increases as problem sources become more difficult, even though models actually perform worse on harder problems. Moreover, worse models can have similar or higher BLEU scores}'' \cite{r24}. Chen et al. also pointed out that similarity metrics like BLEU score may not be a reliable indicator of functional correctness \cite{r19}. They also provided evidence that functionally inequivalent codes often have high BLEU scores. Ren et al. revised the BLEU metric and proposed CodeBLEU \cite{r27} to address the problem that the BLEU metric does not take the features of programming language syntax into consideration. However, the heart of the problem is that a coding problem can have many functionally equivalent solutions that are syntactically dissimilar to each other. In \cite{r31}, as a supplement to test correctness criterion discussed below, Lai et al. used a much-relaxed form of similarity metric called \emph{surface-form constraints}, where each constraint requires the presence/absence of certain specific APIs and/or the keywords to occur in the solution code. 

An alternative to the use of similarity metrics is to test the correctness of generated code. Borrowing the ideas of test-driving software development methodology, Kulal et al. proposed ``pass test cases'' as the correctness criterion to emphasize the importance of functional correctness over text similarity. Each code generation problem is therefore associated with a set of test cases. Du et al. further requires that the set of test cases to be highly adequate and uses test coverages of the reference solutions as measures of the quality of the dataset \cite{r28}. In \cite{r29}, Liu et al. pointed out that the existing benchmark and the approach to evaluating code generation capability suffers from two problems. First, the test cases included in the task as the standard of correctness are insufficient to be a reliable criterion of correctness. Second, the natural language specifications of the tasks are often not precise and complete enough for LLM to process. Their solution to the problems is to generate test cases automatically and check the correctness of LLM output code against the output from the ground-truth solutions. 

The second milestone that Kuala et al. set is now known as the $pass@k$ metric. Kulal et al. asked the SPoC model to produce 100 solutions and considered the task of code generation is successfully completed if one of these solutions is correct in the terms of passing all test cases. This led to the $pass@k$ performance metrics, i.e., probability of completing tasks in $k$ trials. However, a straightforward calculation of $pass@k$ tends to produce large variance \cite{r19}. This problem is addressed by Chen et al. in the evaluation of Codex \cite{r19}. They asked the model to produce a large number $n$ of solutions on each problem, where $n=200$, and counted the number $c$ of solutions that pass the correctness test. They devised a formula to calculate an unbiased estimator of the $pass@k$ metric from the values of $n$ and $c$ for $k < n$. 

The APPS dataset was extracted from open-access websites where software developers share coding problems with each other, including Codewars, AtCoder, Kattis, and Codeforces. Problems that were posed as natural language specifications of what should be coded are manually polished and refined. Duplications were removed using tf-idf features with SVD dimensionality reduction and cosine similarity. Each problem is associated with test cases for checking correctness and ground-truth solutions written by humans. The dataset contains 10,000 problems, with 131,777 test cases and 232,421 ground-truth solutions. The dataset is split into three subsets according to the difficulty levels: Introductory (3,639 problems), Interview (5,000 problems), and Competition (1,361 problems). 

Hendrycks et al. used APPS to evaluate several GPT models. They used two performance metrics: test case coverage and strict accuracy. The former is the average proportions of test cases on which the generated solution passes the test. Therefore, it is better to be called \emph{average pass rate}. The latter is the ratio of the problems that the generated solution passes all test cases. They calculated the overall performance of GPT models using $pass@1$ and $pass@5$. 

In the evaluation of Codex, Chen et al. constructed the dataset HumanEval \cite{r19}. Codex is a LLM that empowers GitHub Copilot for code generation. It is obtained by fine tuning the GPT model using data publicly available from GitHub. The HumanEval dataset contains 164 handwritten problems. Each problem contains:
\begin{itemize}
\item a signature, which defines the syntax format of the function to be generated. 
\item a docstring, which is a text in natural language that describes the function of the code to be generated.  
\item several unit tests, which is used to check the correctness of the code generated. 
\end{itemize}

Chen et al. also used ``pass all test cases'' as the correctness criterion. In other words, a coding solution is regarded as satisfactory only if it passes all the test cases. The signature element of a problem in the HumanEval dataset enables the solution to be executed automatically on test cases to check the correctness of the generated code. It limits the format of the code to be generated. Thus, the dataset can only be applied to generate code at method level. The dataset has an average of 7.7 tests per problem. 

Du et al.'s ClassEval dataset \cite{r28} is a benchmark for generating code at class level while APPS and HumanEval are benchmarks for code generation at standalone function or method level. Similar to APPS and HumanEval, each coding task in ClassEval comprises three elements: a description for the target class, a set of test cases for verifying the correctness of the generated code, and a canonical solution that acts as a reference solution. The description of the target class is a class skeleton with signatures of the methods of the class annotated with natural language texts to specify the functional requirements of the target class and methods and to describe the meanings of the parameters. The ClassEval dataset contains 100 problems to generate 100 classes with 412 methods. The data are manually curated from real-world projects and existing data from HumanEval and MBPP. An important feature of ClassEval is that they emphasized the adequacy of the test cases for the problem to facilitate reliable correctness checking of the generated code. On average, ClassEval's test suites are of $98.2\%$ branch coverage and $99.75\%$ statement coverage. 

Yu et al.'s benchmark CoderEval \cite{r30} aims to provide a wider range of code generation tasks. They classify code generation tasks into six levels: self-contained, slib runnable, plib runnable, class runnable, file runnable, and project runnable. According to Yu et al., APPS and HumanEval are at the first two levels. The dataset contains 230 tasks for Java and 230 for Python code generation in total. 

Lai et al.'s DS-1000 is a code generation benchmark specialized for data science problems in Python\cite{r31}. It contains 1000 coding problems collected from Stack Overflow to reflect natural or real-world coding problems. The problems are also manually modified to create executable context, add reference solutions and additional test cases, and rewrite LLM unreadable data such as figures, etc. Their evaluation of LMMs employed test correctness as well as surface-form constraints that require the presence/absence of certain specific APIs as the keywords. 

\subsection{Analysis of Existing Work}

A common feature of existing work on the evaluation of LLMs is that a single score is used to represent a LLM's performance on one type of tasks such as mathematical reasoning, text comprehension, and code generation, etc. The score is calculated from testing the ML model on a benchmark dataset using one metric. The most important advantage of the approach is that it offers a means of relatively objective comparison of different ML techniques and models, and sometimes with human beings. Such benchmarking has played a significant role in the research on ML techniques as pointed out by Martínez-Plumed et al.\cite{r32}. 

However, the evaluations of LLMs have not reflected their usability as intelligent tools from the users' point of view. For example, ChatGPT is popular to many users while its evaluation scores seem poor \cite{r33}. Codex only achieved a score of $28.8\%$ on $pass@100$\cite{r19} while many users have highly appraised its coding capability. While human testers are in favour of ChatGPT's output on text summary tasks over other state of art ML models, its evaluation score based on ROUGE metric shows the other way\cite{r13}. For the following reasons, such evaluation results may not reflect the usability of the models in real usage scenarios.

First, to be a fair reflection of the performance of a ML model in a specific usage, the dataset used in the evaluation must have the same distribution as the users' input profile. From the constructions of existing datasets, it is hard to see how these datasets represent user input profiles. 

Second, the users of a ML model may have different use scenarios. The input profiles in different use scenarios have different distributions. For example, a senior software developer may have more complicated code generation tasks while junior programmers usually have less difficult coding tasks. A front-end developer may be more likely to produce GUI program code, while a back-end developer would have more tasks to write database related code, and a data analyst would need to generate more statistical data processing code. Therefore, their input profile will have different distributions. Consequently, one ML model having a high performance for one user may perform poorly from another user's point of view if their use scenarios are very different. Even if a dataset that well represents the overall usage of the LLM for the average distribution of input profiles by all users, the figure obtained from such a benchmark will not reflect the usability for a specific type of users. It is desirable to know the performances of the LLM for different types of users in different scenarios. 

Moreover, for many natural language processing tasks, such as to write poems and to engage in dialogues with a human user, there is no objective correctness criterion. A solution that is good enough for one user could be far from satisfactory from another user's point of view. The evaluation of ML models on such types of tasks is inevitably subjective. Using one score to represent the performance of a ML model can hardly represent all users' assessment. 

The above problems also occur in the evaluations of a LLM's capability of code generation. However, there are also a few issues specific to code generation. 

First, the quality attributes evaluated in the existing work are mostly focused on functional correctness of the code. There is a wide range of other important quality attributes because code generation is only a part of the software engineering process. Other quality attributes, like readability, well structuredness, logic clarity, etc., are important for software maintenance, evolution, and reuse. They should also be assessed. Such quality attributes are also particularly important for intelligent interactive coding environments based on reinforcement learning such as Yang et al.'s recent work on InterCode \cite{r34}. 

Second, a piece of code generated by a LLM is often classified in a binary way, i.e., the code is either correct or incorrect. In the real uses of LLMs for code generation, developers mostly use the generated codes as bases for further manual revision \cite{r9, r10, r11}. An incorrect code can also be useful if it can be easily revised into a good solution to the problem. It is desirable to evaluate the performance taking into consideration how close are the generated codes to the required results and how easy to revise the generated code to make a good solution. 

Moreover, the testing and evaluation process can be characterised by ``one attempt per problem''. That is, on each coding problem, the LLM is invoked with one input to generate one or many solutions, no matter whether the input is a one-shot or multiple shot query \cite{r35}. This process is reflected in the $pass@k$ metric. Even if $k > 1$, it is still ``one attempt'' in the sense that it queries the model with only one input, although it tries to get multiple solutions, or equivalently, gives the model multiple chances to answer the same query. However, in the real uses of LLMs as a code generation tool, the developers make many attempts by modifying their inputs until they are satisfied with the output, or they give up after too many attempts. It is desirable to evaluate LLMs' performance on how many such attempts one could expect until a solution good enough is generated. 

\section{The Proposed Methodology}\label{sec:ProposedMethodology}

To address the above problems, we propose a user centric approach to the evaluation of LLMs as intelligent coding tools. 

The general principle of the approach is that, first, the test dataset should be constructed according to the use scenario in which the evaluation is performed. Second, the process of testing and evaluation on each test case should resemble the process in which the ML model is used as a tool. Moreover, the outputs from the ML model should be evaluated on quality attributes that reflect the usability of the output from the users' point of view. Finally, the performance metrics should represent the user's experience with using the ML model. By doing so, we put the user at the center of the evaluation, thus the name of \emph{user centric evaluation}. 

In this section, we will outline our methodology that realises these principles. It will then be delineated and explained in the case study with the evaluation of ChatGPT on its usability of generating R program code.  

\subsection{Multi-attempt Testing Process}\label{subsec:MultiAttemptProcess}
 
In the testing of a LLM, we require the testers to try their best to complete the specified task with as few attempts as possible. Each attempt consists of formulating the best text inputs to invoke the LLM and then inspecting the solution generated by the LLM. If the output from the LLM is unsatisfactory, a second attempt is made by amending the input. The process continues until either the number of attempts reaches the maximum of $k$ of allowed attempts, or a satisfactory solution is obtained. This leads to two performance metrics: \emph{number of attempts}, denoted by $\#attempt_k$, and \emph{time used to complete the test tasks}, or \emph{completion time} for short. They reflect the user experiences in the use of the LLM. This testing process better matches the process that a user typically uses a LLM to generate code than the single-attempt / multi-trial approach and the corresponding performance metric $pass@k$. 

It is worth noting that being user centric, whether a solution is satisfactory should be judged by the user. In our case study reported in Section \ref{sec:CaseStudy} and \ref{sec:ExperimentResults}, a guideline is given to the tester that a solution should be judged as satisfactory if it is correct, or very close to the reference solution. A functionally incorrect solution may well be regarded as satisfactory if a minor editing of the machine generated solution will work. For this reason, when the number of attempts reached the maximum threshold, the test should not be considered as a complete failure. Instead, the quality of the result at the final attempt was also assessed in the case study. On the other hand, even if a generated solution is functionally correct, it may not be regarded as satisfactory, because to be useful it should also meet other quality requirements. which will be discussed in subsection \ref{subsec:QualityAttributes}. 

\subsection{Benchmark with Metadata}

For an evaluation of a LLM to faithfully reflect its users' usage, the test dataset must represent user's uses of the LLM. Our methodology achieves this purpose by constructing a benchmark that supports scenario-based testing in which the LLMs are tested on a series of subsets of the benchmark that each subset represents a specific scenario of using the system. Evaluating a ML model's performance on each subset separately can provide the detailed performance information for different usages of the system. As reported in \cite{r36}, this will also enable us to identify the strength and weakness of the model in different usage scenarios, thus providing directions for future improvement of the model. However, scenario combinations may cause duplicated running of test cases. To support efficient scenario-based testing and evaluating a ML model, we propose to associate metadata to each test case in the benchmark to describe the usage scenarios that the test case represents. This can eliminate the need for duplicated execution of test cases and form subsets for different scenarios easily. This is demonstrated in our case study by combining the scenarios of different difficult levels with the scenarios of generating different types of code. 

\subsection{Quality Attributes for Assessing Usability} \label{subsec:QualityAttributes}

In the real use of a LLM for code generation, a solution produced by a LLM will be examined to determine if it is satisfactory and possibly revised by the human user after accepting the suggested solution. Therefore, the evaluation of a LLM should not only assess the correctness of the generated code, but also its usability. Here, we propose to define a set of quality attributes for this purpose. These quality attributes should be not only on the generated program code but also on the generated texts that explain the code. They should reflect various aspects of examination and revision of the generated solution, such as how easy to understand the generated solution, how easy to revise into a usable solution, and how good the output is in the context of code maintenance, debugging, evolution and reuse. 

For example, the usability attributes used in our case study of ChatGPT on its capability of generating R program code is summarised in Table \ref{tab:QualityAttributes}. 
For each of these quality attributes, the output from the LLM is assessed on the scale from 1 to 5, where 1 is the poorest and 5 the best. Although such quality attributes heavily rely on the tester's subjective assessment, they better reflect users' true opinions on the usability of the tool. 

\begin{table}
\begin{center}
\caption{Attributes of Usability Used in Case Study}\label{tab:QualityAttributes}
\begin{tabular}{|p{1.7cm}|p{6cm}|}
\hline 
\textbf{Attribute} & \textbf{Meanings} \\ 
\hline 
Accuracy & How close the generated code to be correct \\ 
\hline 
Completeness & How complete the generated solution solves the programming problem \\ 
\hline
Conciseness & How concise the generated solution is\\
\hline 
Logic clarity & How clear the code and explanation are presented with a clear and logical flow of thought \\ 
\hline
Readability & How easy to read and understand the generated solution \\ 
\hline 
Structuredness & How well the code is structured \\ 
\hline
Coverage of parameters & How well the parameters provided in the coding problem are used and covered in the solution\\
\hline 
Depth of explanations & How thorough the code is explained by the text in the solution\\ 
\hline
\end{tabular} 
\end{center}
\end{table}

\section{Case Study}\label{sec:CaseStudy}

In this section, we delineate the proposed method by a case study with the evaluation of ChatGPT on its usability as a code generation tool for programming in the R language. We will present the research questions addressed, the design of the experiments, and the construction of a benchmark dataset. The results will be presented in Section \ref{sec:ExperimentResults}. 

\subsection{Research Questions}

Our goal of the experiment is to evaluate ChatGPT's usability as a tool of generating program code in R language. The following are the specific research questions. 
\begin{itemize}
\item RQ0: How good is the usability of ChatGPT as a R program code generation tool? Here, usability is measured in terms of (a) the quality of the generated code as measured on various usability attributes, (b) the user's experiences in using ChatGPT as measured by the average number of attempts, and the average task completion time. 
\item RQ1: How good is the usability of ChatGPT when it is used to solve coding problems of different levels of difficulty? 
\item RQ2: How good is the usability of ChatGPT when it is used to generate different types of program code? 
\end{itemize}

\subsection{Design of the Experiment}

To answer these research questions, we first constructed a benchmark by extracting data from various textbooks on R programming and associated metadata about the source, type of code to be generated, and level of difficulty to each test case in the benchmark. Then, five disjoint test subsets $T_0, T_1, \cdots, T_4$ were drawn at random from the benchmark. These test datasets are of equal size that each consists of 20 test cases. Since these test sets are drawn at random, each of them contains various types of questions and various levels of difficulty, and from different sources. 

The experiments are conducted with the same subject (i.e., a human tester) who is a master's degree student studying on a data analytics program but has no experiences in the use of ChatGPT before the experiment. The testing of ChatGPT is performed sequentially from $T_0$ to $T_4$. 

In the experiment, each test case is a programming task to be completed by the tester via using ChatGPT. The tester is required to formulate a query to invoke ChatGPT with the coding question. The output from ChatGPT is recorded and checked. If the output from ChatGPT is not satisfactory in terms of the correctness of the generated code, a further prompt is submitted to ChatGPT. This is repeated until a satisfactory output is obtained or it is terminated after 10 cycles of input/output. Each cycle is called an \emph{attempt}. The time used to complete the task on each test case and the number of attempts engaged with ChatGPT were also recorded. The final output is then manually assessed on the quality attributes presented in Table \ref{tab:QualityAttributes}. 

Statistical analysis is conducted on the union $T$ of these test sets, i.e., $T = T_0 \cup T_1 \cup \cdots \cup T_4$, for the overall performance of ChatGPT to answer research question RQ0. The test set $T$ is then split into subsets according to difficulty levels and types of program codes to answer research questions RQ1 and RQ2 using the metadata associated with the test cases without re-executing them and re-assessment of quality. To answer research question RQ3, the trends of performances changes are analysed separately on the disjoint subsets $T_0, T_1, \cdots, T_4$. 

\subsection{Construction of Dataset}

As far as we know, there is no existing dataset available for generating R program codes. Being a special purpose language targeting data analysis and statistics, R contains many special features. General datasets of code generation and those for other languages cannot provide a fair evaluation. Therefore, we constructed a dataset specifically for the generation of R program code. 

\subsubsection{Source of Data}
	
The data are gathered from exercise questions of accredited R programming textbooks, which covers a wide range of topics, including uses of various language facilities and libraries, numerical calculations, data structural manipulations and transformations, data visualization and statistical analysis, etc. The questions are of different difficulty levels. These questions are well presented since they are from well written textbooks. These books are primarily used in the IT industry to educate and train professional programmers who have no experience of programming in the R language. We have collected all questions from six textbooks given in references \cite{r37,r38,r39,r40,r41,r42}. They are referred to as B1 to B6, respectively, in the sequel. 

Each test case in the benchmark contains the exercise question as the programming task and the answer provided by the textbook as the reference solution of the task. 

\subsubsection{Metadata}

In addition to the question and the reference solution, we also contain the following metadata to each question in order to support scenario-based testing. 

\textbf{(a) Difficulty Levels}. For each programming task in the benchmark, a difficulty level is assigned manually according to the following criterion. 

\begin{itemize}
\item \emph{Easy}: when the program code to be generated is essentially a simple invocation of a library function. 
\item \emph{Medium}: when the program code to be generated is beyond a simple invocation of predefined functions. However, it does not involve complicated data structure and algorithms. 
\item \emph{Hard}: when the program code to be generated requires complex data structures and algorithms. 
\end{itemize}

\textbf{(b) Types of Questions}. For each question in the benchmark, a type is assigned according to the type of code to be generated. We classify coding questions into the following types. 
\begin{itemize}
\item \emph{Numerical}: The task is to generate a numerical calculation program for a given calculation formula or equation. 
\item \emph{Statistical}: The task is to generate a program that performs statistical analysis of a dataset, such as a code to do linear regression. 
\item \emph{Structural}: The task is to change the structure of data, for example, to remove or add a feature from/to the elements of a dataset. 
\item \emph{Programming}: The task is to use a particular language facility in problem solving, such as to use a loop statement, input/output statement, etc. 
\item \emph{Visualization}: The task is to generate a code to visualize a dataset, such as to create a plotting graph, etc. 
\item \emph{Exploratory}: The task is to generate code that performs exploratory analysis of a given dataset. 
\end{itemize}

\textbf{(c) Sources of the data}. The source of a question is also provided as a metadata of the test case. Since our questions are all from textbooks, the metadata contains the title of the book, the year of publication and the chapter of the book where the original question is in. This metadata is potentially useful to analyse how well a LLM deals with new programming tasks and uses new libraries and facilities of the R programming language, although it is not used in the experiments reported in this paper. 

The metadata are represented in JSON format. Figure 1 shows the structure of the JSON representation of test cases. An example is given in Figure 2. 

\begin{figure}[htbp]
\centerline{\includegraphics{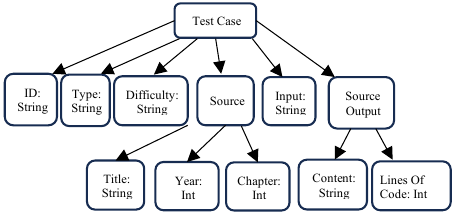}}
\caption{Structure of JSON Representation of Test Cases.}
\label{fig1}
\end{figure}

\begin{figure}[htbp]
\centerline{\includegraphics{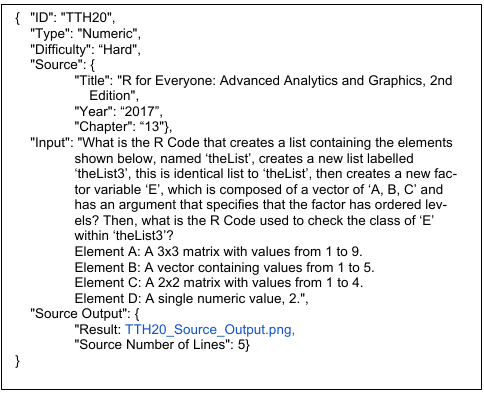}}
\caption{Example of test case represented in the JSON format.}
\label{fig2}
\end{figure}

The dataset contains 351 test cases in total. Table 1 shows the statistical characteristics of the dataset. Each row of the table gives the number of test cases in the benchmark, where the first column is the feature, columns B1 to B6 are the numbers from different sources, and the total number in the column Total. The final row of the table shows the total number of test cases from each source. 

\begin{table}[htbp]
\caption{Statistical Characteristics of the Dataset}
\begin{center}
\begin{tabular}{|l|c|c|c|c|c|c|c|}
\hline
\textbf{Feature} &\textbf{B1}&\textbf{B2}&\textbf{B3}&\textbf{B4}&\textbf{B5}&\textbf{B6}&\textbf{Total}\\
\hline
Easy&26&32&19&12&27&14&130\\
\hline
Medium&29&36&15&11&25&10&126\\
\hline
Hard&25&28&10&10&13&9&95\\
\hline\hline
Numerical&10&14&8&9&13&8&62\\
\hline
Statistical&12&15&6&4&17&6&60\\
\hline
Structural&16&17&10&5&9&9&66\\
\hline
Programming&20&22&12&8&7&4&73\\
\hline
Visualization&16&18&8&4&15&4&65\\
\hline
Exploratory &6&10&8&3&4&2&33\\
\hline\hline
\textbf{Total}&\textbf{80}&\textbf{96}&\textbf{44}&\textbf{33}&\textbf{65}&\textbf{33}&\textbf{351}\\
\hline
\end{tabular}
\label{tab1}
\end{center}
\end{table}

\section{Experiment Results} \label{sec:ExperimentResults}

This section reports the experiment results on the overall performance of ChatGPT on generating R program code, and the performances in various scenarios in which different types of code are to be generated, and the impacts of difficulty level on performance.  Finally, the potential threats to the validity and generalisability of the experiment results will be discussed. 

\subsection{Overall Performance}

\begin{figure}[htbp]
\centerline{\includegraphics{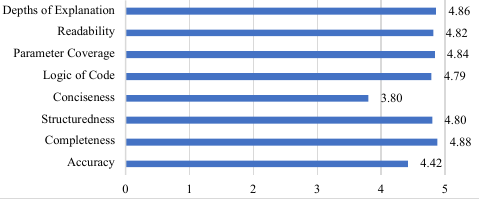}}
\caption{Average Performance Scores on Quality Criteria.}
\label{fig3}
\end{figure}

To answer research question RQ0, the overall performance of ChatGPT on its usability as a R code generator is analysed on all test cases in the test set $T=T_0 \cup \cdots \cup T_4$. The experiment results are shown in Figure \ref{fig3} with the average performance scores on various quality criteria. 
 
As the data shows, on average ChatGPT performed very well on all quality criteria except conciseness, which has an overall average score of 3.80 out of 5, i.e., 76\%. The second weakest quality aspect is accuracy, which has an average score of 4.42 out of 5, i.e. 88.4\%. 

\begin{figure}[htbp]
\centerline{\includegraphics{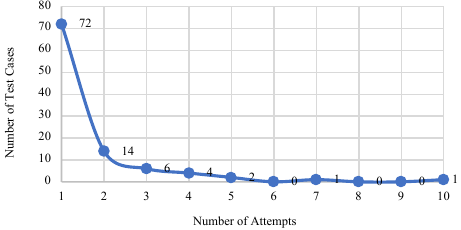}}
\caption{Distribution of the Numbers of Attempts.}
\label{fig4}
\end{figure}

ChatGPT also performed very well on objective metrics, i.e. the number of attempts and the time to complete the task. Figure \ref{fig4} shows the distribution of the numbers of attempts. The average number of attempts is only 1.61. Among 100 test cases, 72\% is completed with one attempt, 14 with 2 attempts, 6 requires 3 attempts. On 98\% of the test cases, ChatGPT successfully generates accurate results with less than or equal to 5 attempts. Only 2 test cases required more than 5 attempts, where one completed the task with 7 attempts, while on only one test case it failed to generate a satisfactory result after 10 attempts. 

\begin{figure}[htbp]
\centerline{\includegraphics{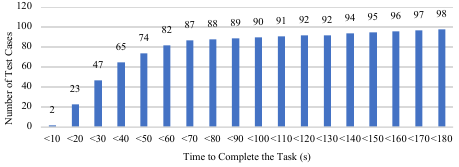}}
\caption{Distribution of Completion Times (Seconds).}
\label{fig5}
\end{figure}

The time to complete a task is mostly spent on formulating text inputs as the queries to the ChatGPT and reviewing the output obtained from ChatGPT. ChatGPT responds to queries so fast that the time is negligible. Figure \ref{fig5} shows the distribution of the time needed to complete the task of generating R code, where X-axis is the time required to complete the task, Y-axis is the number of test cases. The average of completion time is only 47 seconds while the median is 31.50 seconds. It is worth noting that 90\% of test cases can be completed within 100 seconds, while 98\% of test cases were completed within two and half minutes. The two cases that required more than 5 attempts took more than 180 seconds, where the one that was given up after 10 attempts took 462 seconds. 
 
The objective metrics of performance confirmed our conclusion that ChatGPT’s usability as a code generation tool is satisfactory in general. 

\subsection{Performance of Generating Different Types of Code}

\begin{figure}[htbp]
\centerline{\includegraphics{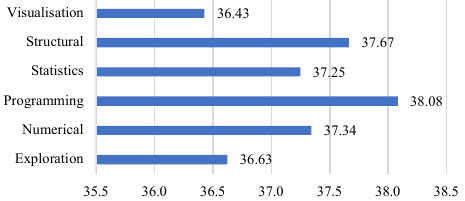}}
\caption{Average Total Quality Scores of Various Types of Questions}
\label{fig6}
\end{figure}

A detailed analysis of ChatGPT's performance on generating different types of R code reveals that it has the best performance on generating general programming type of code with an average total quality score of 38.08 out of 40 (i.e. 95.2\%), while weakest in generating visualization and exploration types of code with the average total quality scores of 36.43 (91.1\%) and 36.63 (91.6\%), respectively. The results are depicted in Figure \ref{fig6}. 

\begin{figure}[htbp]
\centerline{\includegraphics{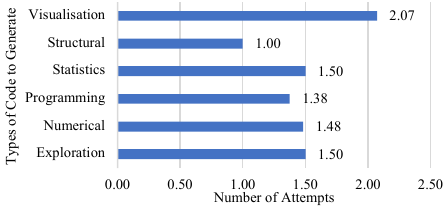}}
\caption{Average Numbers of Attempts for Different Types of Code}
\label{fig7}
\end{figure}

\begin{figure}[htbp]
\centerline{\includegraphics{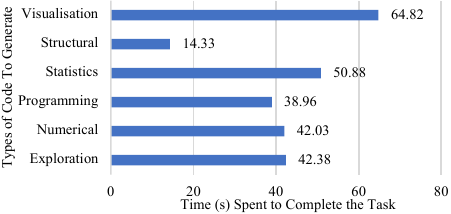}}
\caption{Average Completion Times for Different Types of Code}
\label{fig8}
\end{figure}

The observation that ChatGPT is weakest on the generation of visualization code is confirmed by the analysis on the objective metrics, i.e. the average number of attempts and the average time to complete tasks; see Figure \ref{fig7} and \ref{fig8}.

\begin{figure}[htbp]
\centerline{\includegraphics{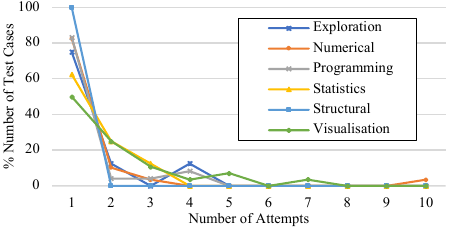}}
\caption{Distributions of The Numbers ofAttempts for Different Types of Code}
\label{fig9}
\end{figure}

\begin{figure}[htbp]
\centerline{\includegraphics{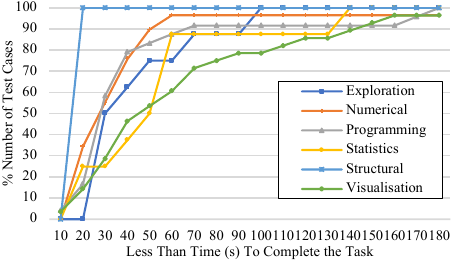}}
\caption{Distributions of Task Completion Times for Different Types of Code}
\label{fig10}
\end{figure}

The distributions of attempts and task completion times are shown in Figure \ref{fig9} and \ref{fig10}, respectively. 

\begin{figure}[h]
\centerline{\includegraphics{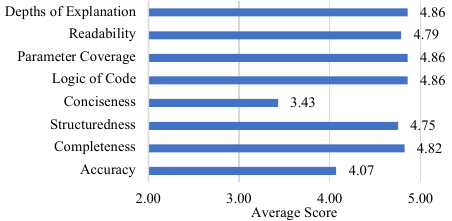}}
\caption{Average Quality Scores on Visualization Questions}
\label{fig11}
\end{figure}

Figure \ref{fig11} shows the average quality scores for visualization type of code on various quality attributes. The data shows that conciseness is the weakest aspect of quality on generating visualization code, which scored 3.43 out of 5, i.e. 66.8\%. 

\subsection{Impacts of Difficulty Levels on Performance}

\begin{figure}[h]
\centerline{\includegraphics{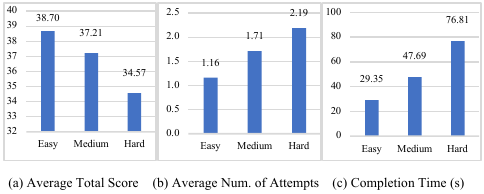}}
\caption{Usability on Different Levels of Difficulty.}
\label{fig12}
\end{figure}

Will ChatGPT's usability decrease as the difficulty of the task increases? The experiment data shows clearly that the average total quality scores on subjective quality criteria decrease as the level of difficulty increases, while the average number of attempts and the average task complete time increase as the difficulty level increases; see Figure \ref{fig12}. 

\begin{figure}[h]
\centerline{\includegraphics{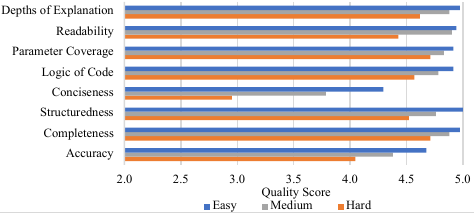}}
\caption{Usability on Various Quality Criteria on Different Difficulty Levels}
\label{fig13}
\end{figure}

Moreover, when measuring usability on each individual quality criterion, the experiment data also clearly demonstrate the monotonicity of performance change with the difficulty level; see Figure \ref{fig13}.

\subsection{Potential Threats to The Validity of Experiment Results}\label{sec:ThreatsToValidity}

The scale of our experiments is small in terms of the number of subjects involved in the testing process due to our resource limitation. This forms a potential threat to the validity and generalizability of the results. However, the feasibility of the proposed testing method is not affected. The potential threat can be eliminated by repeated experiments with more subjects. This suggests a topic for future work to repeat the experiments with more testers. 

The manual assessment on quality attributes is subjective, which depends on the subject's knowledge, experience, and ability. This causes another potential threat to the validity of the results. Efforts have already been made to confine the threat by applying the metrics consistently according to the clear definitions of the metric values. Thus, the risk is minimized. Moreover, our experiment data show that the results obtained from such subjective judgements are consistent with objective metrics of the number of attempts and time to completion. Thus, the results should be trustworthy. For future research, we are working on using more objective metrics about the quality of program code developed in software engineering research, like cyclomatic complexity, coupling and cohesion, code smell, etc., in complementary to the subjective metrics proposed and used in this paper. It is worth further research to find out if machine generated codes are of high quality on these attributes designed for traditional human written code, and whether such metrics can provide meaningful assessment of the usability of machine generated code. 

A factor that has an impact on the experiment results is the criterion to determine whether a solution is satisfactory in the multi-attempt testing process. In principle, being user centric, it must be judged by the human users who engage in the testing. In our experiment, the guideline given to the tester is that the generated code is satisfactory if it is correct when compared with the reference solution or is close enough to the reference solution. In other words, a functionally incorrect solution may well be regarded as satisfactory if a minor editing of the machine generated solution will work. However, in practice, such reference solutions are usually unknown. Therefore, there is a gap between the judgements of satisfactoriness by the testers and the real users. This is a potential threat to the validity of the experiment results. It is desirable to give a more rigorous criterion to judge how satisfactory a generated solution is. However, defining such a rigorous criterion is a non-trivial issue and worth further research. Study of metrics on the usefulness of machine generated solutions could be helpful. 

Our dataset is constructed by extracting exercise questions from textbooks. An advantage is that the questions are well presented in terms of the clarity of the language and coverage of the programming language facilities and usages as well as the correctness of the reference solution. Thus, little manual editing and filtering of data is required while other benchmarks constructed by extracting data from open-source code repositories have required heavy workload on manual editing and cleaning of the data. Consequently, we have managed to construct a relatively large dataset with a small amount of human resource. It also minimizes the possibility of testing LLM on incorrect questions. A potential threat to the validity of the experiment is that questions from textbooks may not represent the coding tasks in the real uses of LLMs. It is worth further research on the validity of the dataset via studying the differences between real input from users and the data in our benchmark. 

We have included three types of metadata to our benchmark data: the type of code to be generated, the difficulty level of the task to be completed, and the source of the data. The first two represent two kinds of scenarios, which can be combined to form various specific usage scenarios as demonstrated in the case study. Since the metadata are manually assigned to the test cases in the dataset, a potential threat to the validity of the experiment results can be caused by errors in the metadata. In the experiment, we have taken mitigation means to minimize the error rate by giving clear definitions of the values of metadata. For example, a coding task is assigned with ``hard'' as its difficulty level, if and only if ``when the program code to be generated requires complex data structures and algorithms.'' In the literature, some datasets do classify the coding tasks into different difficulty levels. For example, APPS dataset has three subsets of data on three difficult levels: Introductory, Interview, and Competition. However, there is no definition of the criterion on how the tasks were classified \cite{r24}. Our approach enables metadata to be assigned systematically and consistently, thus the chances of human errors are reduced. As for future work, it is worth studying other types of metadata and other values of metadata, for example, difficulty levels could be more than three. Another topic for future work is how to check the correctness of metadata. 

\section{Comparison with Existing Works}\label{sec:Comparison}

Existing evaluations of LLM models are mostly conducted through benchmarking on given test datasets. While this enables objective comparisons of LLMs, the results hardly reflect their usability from the user's perspective. The method proposed in this paper is a scenario-based approach. It aims at evaluation of a LLM on a number of scenarios that each scenario is represented by a subset of test cases. Understanding the performances of a LLM  on various different scenarios is particularly important when deciding on whether to use a LLM in a specific use scenario. The difficulty is that there is a wide range potential uses of LLMs, while testing and evaluating a LLM is expensive and difficult. Our solution is to assign metadata to each test case in the benchmark dataset to describe various aspects of its usages. This enables flexible tailoring of the dataset to generate subsets that represent various use scenarios, and to adjust the evaluation on different scenarios with existing test results without re-run test cases. It is different from existing benchmarks for evaluating code generation capability. As discussed in Section \ref{sec:RelatedWork}, these benchmarks only contain the coding task, the reference solution(s) and/or a set of test data for checking the correctness of the machine generated solution. 

Our method also differs from existing work on the testing process. Our multi-attempt process of testing on each test case mimics the real uses of LLMs. In this process, the tester tries to complete the testing task through a series of attempts until a satisfactory solution is obtained or gives up after a fixed number of maximal allowed attempts. In each attempt, the tester formulates an input based on the previous results and the task to be completed, invokes the ML model, and then inspects the output to decide whether the solution is satisfactory. As discussed in Section \ref{sec:RelatedWork}, this contrasts with the multi-trial/single-attempt process currently used by the existing work on evaluating code generation. In the multi-trial process, a number of solutions are generated for the same input. Our process is closer to the real uses of LLMs. 

Moreover, in our method, the quality of an output is measured by a set of metrics on the usefulness of the generated code from the users' perspective. This contrasts with existing works on evaluations of code generation, where the functional equivalence of a machine produced solution to the reference solution(s) are used as the sole judgement on the quality of the code generated. In the real uses of LLMs for code generation, a piece of code is considered as useful even if it is not functionally correct. In many cases, a skeleton of code is acceptable even if it cannot be compiled. Our quality attributes focus on the usability of a solution via considering its readability, logic clarity, well-structuredness, etc. Thus, again the method is closer to the real uses of LLMs. 

Finally, we define two metrics to measure user experiences in the uses of a LLM: (1) $\#attempt_k$, i.e. the average number of attempts, and (2) the average task complete time. They differ from the $pass@k$ metric, which is currently widely used in evaluation of code generation capability. Our metrics better reflect the user experiences in the use of the LLM. 

Therefore, our proposed approach to the evaluation of LLM on code generation can be characterized as user centric for its focus on user experiences in the uses of the LLM. This is reflected in all aspects of the testing and evaluation, including the information contained in the benchmark dataset, the testing process, the measurement of solution quality and the performance metrics. 

Our case study results are consistent with the good user experiences and general opinions on ChatGPT. We have also identified the weakness as well as the strength of the LLM model. 

\section{Conclusion}\label{sec:Conclusion}

This paper proposed a user centric methodology to the testing and evaluation of ML models for code generation. The main contributions are as follows. 

First, we introduced the uses of metadata assigned to test cases in a benchmark to support flexible construction of test datasets that represent various usage scenarios of ML models and flexible analysis of testing results. 

Second, we proposed the multi-attempt testing process so that the testing on each test case mimics the real uses of a ML model by the user. 

Third, we defined a new the performance metric $\#Attempt_k$, i.e. the \emph{average number of attempts} . Together with the average task completion time, it provides an objective measure of the user experiences in the use of a ML model to generate program code. 

Finally, we employed a set of quality attributes to measure the usability of generated program code and explanation text. 

The proposed methodology covers all aspects of ML model evaluation from the construction of benchmark dataset, testing process, quality attributes to be assessed, to the metrics to evaluate ML model's performance. In comparison with the existing approaches, its main advantage is that the evaluation results can better reflect users' experiences in the use of a ML model. Moreover, by providing evaluation of a ML model on a spectrum of scenarios, the results can provide more insights of the weakness and strengths of the model for developers to improve its performance. 

There are several interesting topics worth research to further develop the methodology. For example, the user centric testing method proposed in this paper is expensive to apply because of human users' involvement in the testing process. It is worth further research to reduce the cost and to improve efficiency through testing tools and automation. 

It is worth noting that evaluating LLM is inevitably subjective due to the nature of NLP. Instead of trying to completely avoid subjective assessment, we minimize the potential bias of the assessment by decomposing the overall subject opinions into a set of quality attributes that represent different aspects of subjective opinions and use Likert's 5 levels of scores to normalize the assessments. This approach has been well developed and widely applied in software engineering and HCI to evaluate software quality. How to further improve the methodological rigour of user centric evaluations of LLMs will be of great importance.  

Finally, a LLM's capability of generating program code is not only useful to software development as a coding tool, but also for many other purposes such as education. For example, a student could learn coding with a LLM by raising questions in natural language and getting answers from the LLM with program codes and detailed explanations. Another possibility is to help students to understand example program code by inputting a piece of code to a LLM and getting an explanation of the code. Moreover, a student's answer to an exercise or examination question could be checked and marked by a LLM for its correctness and producing a feedback to the student. If a student's answer to an exercise question is incorrect, the LLM could give the reasons why it is incorrect and delineate on how to correct it. It is worth further research to develop a methodology for the evaluation of LLMs in such uses based on the user centric principle.  

\section*{Acknowledgment}

The work reported in this paper is funded by the 2020 Research Excellence Award of Oxford Brookes University.

\end{document}